\documentclass[onecolumn,showpacs,preprint,amsmath,amssymb]{revtex4}
\usepackage{graphicx}
\usepackage{dcolumn}
\usepackage{bm}

\begin{document}

\title{Microwave distribution in stacked Bi$_{2}$Sr$_{2}$CaCu$_{2}$O$_{8+x}$
intrinsic Josephson junctions in a transmission-line geometry}
\author{Myung-Ho Bae}
\author{Hu-Jong Lee}
\email{hjlee@postech.ac.kr} \affiliation{Department of Physics,
Pohang University of Science and Technology, Pohang 790-784,
Republic of Korea }
\author{Jinhee Kim}
\author{Kyu-Tae Kim}
\affiliation{Electricity Group, Korea Research Institute of
Standards and Science, Daejon 305-600, Republic of Korea}

\begin{abstract}
The microwave distribution inside a rectangular stack (15
$\mu$m$\times$0.72 $\mu$m$\times$60 nm) of
Bi$_{2}$Sr$_{2}$CaCu$_{2}$O$_{8+x}$ intrinsic Josephson junctions
(IJJs) was studied. The stack was microfabricated into a
transmission-line geometry, with a-few-hundred-nm-thick Au layers
deposited on top and bottom of the stack. The microwave
distribution was monitored by measuring the anomalous suppression
of the tunneling critical current of the IJJs with varied
microwave power at frequencies in the W band. This technique can
provide valuable information on the microwave transmission modes
inside the sandwiched stack of IJJs, which is utterly important
for the high-frequency device applications using IJJs, such as
fluxon-flow THz oscillators.
\end{abstract}


\maketitle

The microwave response of intrinsic Josephson junctions (IJJs)
formed in Bi$_{2}$Sr$_{2}$CaCu$_{2}$O$_{8+x}$ (Bi-2212) single
crystals has been studied extensively
\cite{KleinerMuller,Wang1,Doh,Irie,Wang2} since the observation of
the Josephson effects in the material \cite{Kleiner}. The
intrinsic nonlinear character of Josephson junctions and the large
gap value in CuO$_2$ superconducting bilayers make IJJs a
promising candidate for microwave applications. There have been
successful observation of Shapiro steps \cite{Wang1}, coherent
mode splitting \cite{Doh}, phase-locked steps of fluxon-induced
microwaves \cite{Irie}, and harmonic frequency mixing
\cite{Wang2}, in the frequency range of microwaves. Coherent
radiation from resonances between plasma modes and field-generated
fluxons in stacks of IJJs has been predicted theoretically
\cite{Machida}. Non-Josephson-like radiation caused by the
fluxon-flow motion has been observed in stacks of IJJs
\cite{Hechtfischer}.

Recently, quasi-one dimensional mesa structures where the width of
the mesa was comparable to the Josephson penetration depth
$\lambda_{J}$ were employed for investigating the coherent fluxon
motion in a dc magnetic field. Consequent fluxon-flow steps in the
Josephson state were observed \cite{Heim}. In order to realize a
local fluxon-flow radiation device using IJJs, however, careful
characterization is required for the propagation behavior of the
induced microwaves in a stack of IJJs with well-defined geometry.
Recently, good coupling of THz-range microwaves to a stack of
Bi-2212 IJJs has been successfully achieved by employing a
bow-tie-shape transmission-line geometry \cite{Wang1}.

For detailed investigation of the distribution of microwaves in
the transmission-line geometry of IJJs, we fabricated a
rectangular-shape stack of IJJs sandwiched between Au films at
bottom (100 nm) and top (400 nm) of the stack. Microwaves of
varied frequencies (in the range of 70-95 GHz) and power were
irradiated on the rectangular stack. We observed anomalous
microwave-induced suppression of the tunneling critical current of
stacked IJJs. The results were consistent with standing mode
distribution of the microwaves along the $c$ axis, with the stack
of IJJs as an effective dielectric medium.


Long quasi-one-dimensional stacks were prepared in the following
way. Bi-2212 single crystals were grown by the
solid-state-reaction method \cite{Kim}. A single crystal was first
glued on a sapphire substrate using negative photoresist (OMR-83)
and was cleaved until an optically clean surface was obtained.
Then a 100-nm-thick Au film was thermally deposited on top of the
crystal to protect the surface. A mesa with the size of $\sim$25
$\mu$m$\times$0.72 $\mu$m$\times\sim$60 nm was then prepared by
photolithographic patterning and Ar-ion-beam etching [Fig. 1(a)].
The surface of the patterned mesa was fixed to another sapphire
substrate using negative photoresist and the basal part of the
mesa was cleaved away. A 100-nm-thick Au film was again deposited
on this freshly cleaved crystal surface, leaving a stack of IJJs
sandwiched between two Au electrodes [Fig. 1(b)]. A
few-$\mu$m-long portion on both ends of the stack was etched away
subsequently to get the bottom Au electrode exposed for $c$-axis
transport measurements [Fig. 1(c)]. Finally, 300-nm-thick
Au-extension pads were attached [Fig. 1(d)]. This double-side
cleaving \cite{Wang1} process allows one to prepare a thin stack
of IJJs with electrodes on top and bottom, {\it without the basal
part}. The resulting structure enables one to effectively prevent
the fluxons in the basal stack from interfering with the fluxon
motion in the mesa. It also provides a transmission line for
microwaves so that one can conveniently tune their propagation
conditions in the line.

The microwave from a Gunn diode was transmitted through a
waveguide and coupled inductively to the stack. The measurements
were made in a two-terminal configuration at $T$=4.2 K [Figs. 1(d)
and 1(e)] while sweeping the bias current at a several tens of Hz.

Figs. 2(a) and 2(b) exhibit the evolution of the current-voltage
({\it I-V}) characteristics of our stack in the microwave of
frequencies 72.3 GHz and 94.4 GHz, respectively, with gradually
increasing the irradiation power. The uppermost sets of curves in
both figures were obtained for extremely weak power. Each branch
in the data is the quasiparticle branch from each IJJ in the
stack. Thus the number of branches indicates that the stack
contained $\sim$40 IJJs, which corresponded to the thickness of
$\sim$60 nm. Two-terminal measurements gave a finite contact
resistance of 1050 ohm from top and bottom contacts manifested as
the finite voltage in the first branch of the data. Branches in
the low-bias region indicate that about eight junctions had
smaller critical current $I_{c}$ than the other junctions,
possibly due to the proximity effect by the Au layers on the
stacks close to the top and the bottom \cite{Kim} or due to any
surface degradation during the fabrication. Thus in the remainder
of this paper these junctions will be identified as 'the surface
junctions'.


The critical current $I_{c}$ of the inner junctions is 0.1 mA and
the normal-state resistance (with the contact resistance
subtracted) $R_{n}$, determined following the fitting procedure as
in Ref. \cite{Won}, is 33 $\Omega$ per junction so that the
characteristic voltage is about 3.3 mV (or equivalently
$\simeq$1.6 THz). The plasma frequency $f_p$ turns out to be 30
GHz, determined with junction capacitance of 8.5 pF, the effective
dielectric constant $\epsilon_r\simeq$100 (extracted from the
quasiparticle return current), and the Josephson penetration depth
$\lambda_{J}$ of 0.34 $\mu$m. The corresponding Swihart velocity
\cite{Doh} is $2\pi f_p\lambda_J$=6.35$\times$10$^4$ m/s. The
order of magnitude of this rough estimate is in reasonable
agreement with the usual value in the mesa structure \cite{Heim}.

The stack of IJJs sandwiched between two Au layers can be
considered as a transmission line for a microwave, where the
Bi-2212 stack consisting of extremely thin CuO$_2$ layers (0.3 nm)
and insulating Sr-O and Bi-O layers (1.2 nm) acts as an effective
dielectric medium as a whole. In the usual mesa structure without
wave-confining electrodes microwaves are distributed rather
uniformly over the mesa and the basal stack as well. In
comparison, specific modes can form along the $c$ axis inside our
sandwiched stack. Since $I_{c}$ of a Josephson junction is
suppressed with microwave irradiation \cite{Doh,Irie} one can
monitor the microwave distribution inside a stack by measuring the
change in $I_{c}$ of IJJs. In this sense, the IJJs are utilized as
intrinsic EM-wave-detecting sensors inside the stack. For $f$=72.3
GHz in Fig. 2, $I_{c}$ of all the junctions decreases
monotonically with increasing the microwave power from -51 dB up
to -3 dB [see Fig. 2(c) also], which is in general agreement with
the microwave response observed previously in a mesa structure
\cite{Doh}. By contrast, for $f$=94.4 GHz, $I_{c}$ of the 3-5
lowest-bias branches from the surface junctions remains almost
unchanged for microwave power up to -29 dB, while $I_{c}$ of the
rest of the junctions decreases monotonically as for $f$=72.3 GHz.
For power of -29 dB, $I_{c}$ of the rest of the branches becomes
even smaller than that of the surface-junction branches. For
$f$=94.4 GHz the $I_{c}$ of the surface-junction branches starts
decreasing only beyond -29 dB.

Figs. 3(a)-3(c) show the suppression of $I_{c}$ of the 3rd
($I_{c}^{3rd}$) and the 9th ($I_{c}^{9th}$) branches for $f$=72.3,
84.2, and 94.4 GHz, respectively, as a function of the square root
of the microwave power. $I_{c}^{9th}$ is suppressed with
irradiation power in a similar manner for all three frequencies.
For $f$=72.3 GHz, $I_{c}^{3rd}$ is suppressed monotonically with
power but with a much slower rate than $I_{c}^{9th}$, merging to
the value of $I_{c}^{9th}$ around $P^{1/2} \sim$0.1 in an
arbitrary unit. For $f$=84.2 GHz, $I_{c}^{3rd}$ is almost
insensitive to the power up to $P^{1/2} \sim$0.1, above which
$I_{c}^{9th}$ becomes even smaller than $I_{c}^{3rd}$. The trend
becomes more conspicuous for $f$=94.4 GHz, where $I_{c}^{3rd}$ is
virtually constant up to $P^{1/2} \sim$0.17, above which $I_{c}$
of the two branches merge. For the power in the range
$\sim0.1<P^{1/2}<\sim$0.17, $I_{c}^{9th}$ again becomes smaller
than $I_{c}^{3rd}$, which is the same phenomenon as discussed in
relation with Fig. 2(b). This strongly indicates that the
microwave distribution along the $c$ axis depends on its
frequency. The microwave was almost uniformly distributed across
the stack for frequencies below $\sim$84 GHz, as evidenced by the
overall suppression of $I_c$ as in Fig. 2(a), but it was more
intense in the middle of the stack for frequencies above $\sim$84
GHz. Fig. 3(d) shows the gradual change in the {\it I-V} data for
a frequency range of 71.7- 94.5 GHz, taken for varied irradiation
power that gives the maximum difference of $I_{c}$ between the 3rd
and the 9th branches for a given frequency. The average values of
$I_{c}$ for the frequency range of 70 GHz and 90 GHz, denoted by
two lines in Fig. 3(d), are about 25 $\mu A$ and 38 $\mu A$.

The observed results can be explained if the sandwiched
rectangular stack behaves as a transmission line \cite{Pozar},
where a microwave is confined as a standing wave only along the
$c$ axis. The mode frequency is expressed as $f_n$=$nc_d/2d$,
where $c_{d}$ is the speed of a microwave in a stack of thickness
$d$ and $n$ ($>$0) is the integer mode number. We assume the
frequency $\sim84.2$ GHz is the lowest mode frequency
$f_c$=$c_d/2d$, because at frequencies below this value a
microwave was uniformly distributed inside our stack presumably
without forming a mode. For frequencies higher than this value the
magnitude of $I_c$ of the two branches were inverted as in Fig.
3(b). This value of $f_c$ gives $c_d$=1.01$\times$10$^4$ m/s. Fig.
1(f) shows the schematic distribution of a microwave intensity
inside the transmission-line stack, which is consistent with the
lowest-mode microwave distribution. For $f>$84.2 GHz, microwaves
can transmit through the inner junctions with a standing-wave mode
along the $c$ axis. The skin depth $\delta_s$ of an Au layer at
$f_c$, estimated with the measured conductivity of an Au layer
1.31$\times$10$^8$ mho/m at 4.2 K, was about 150 nm. Since the
bottom Au layer was marginally thick to confine the microwave some
leak of the microwave out of the transmission line was expected.

One notes that our sandwiched stack is not a simple dielectric
medium, but consists of serially connected IJJs. The microwave
propagation inside stacked IJJs themselves is governed by the
Josephson plasma modes with the mode velocity \cite {Kleiner2},
$c_{n}=c_0 /\sqrt{1-\mbox{cos}[\pi n/(N+1)]}$, where $c_0$ is the
Swihart velocity and $N$ is the number of Josephson junctions in
the stack.  The mode number $n$ varies from $1$ to $N$. The lowest
mode velocity, $c_N$, is approximately given as
$c_0/\sqrt{2}=4.5\times$10$^4$ m/s for our stack. This value turns
out to be the same order as but about four times higher than
$c_d$, so that the microwave propagation in a transmission line,
regarding the stack of IJJs as an insulating medium, is compatible
with the lowest-mode plasma propagation in the stacked IJJs. If
the finite skin depth is taken into account, however, $d$ should
be replaced by an effective value, which will give rise to a
higher value of $c_d$. The magnetic field component of microwaves
of high irradiation power generates Josephson fluxons in long
junction stacks \cite{Doh}, which give the fluxon-flow resistance
as in Fig. 2(c). The observed suppression of $I_{c}$, however,
occurred before the fluxon-flow state was established.

In conclusion, the microwave distribution inside a stack of IJJs
in a transmission-line geometry was investigated by measuring the
suppression of the tunneling critical current of the IJJs with
varied microwave power at frequencies in the W band. We identified
a cut-off frequency for the mode formation. Microwaves were
uniformly distributed below the cut-off frequency, but became
stronger in the middle of the stack above the cut-off frequency.
Thus, to apply a microwave uniformly across an entire stack of
IJJs, it may be desirable to design the the stack geometry so as
to select measurement frequencies ranging between the Josephson
plasma frequency of individual IJJ and the lowest cut-off
frequency of the sandwiched stack. The technique used in this
study will provide valuable information for the device
applications using IJJs in the microwave frequency range.

We wish to thank valuable discussion with and helps from Nam Kim
in KRISS and Young-Soon Bae in POSTECH. This work was supported by
the National Research Laboratory project administrated by KISTEP.

\newpage
\textbf{FIGURE CAPTIONS}
\\

Figure 1. (a)-(d) Schematic illustration of the sample fabrication
procedure using the double-side-cleaving technique, (e)
configuration for microwave application, and (f) the intensity
distribution of microwaves inside a stack along the $c$ axis.
\\
\\
\\
Figure 2. {\it I-V} characteristics of the stack under the
microwave irradiation of frequencies (a) 72.3 GHz and (b) 94.4
GHz, and various irradiation power. (c) {\it I-V} characteristics
in the fluxon-flow state at 72.3 GHz for various power ranging
between -25 dB and -3 dB from top to bottom.
\\
\\
\\
Figure 3. (a) The suppression of $I_{c}$ of the 3rd and the 9th
branches as a function of square root of microwave power. (b)
Gradual change in the {\it I-V} data for a frequency range between
71.7 GHz and 94.5 GHz for varied power that gives the maximum
difference of $I_{c}$ between the 3rd and the 9th branches for a
given frequency. The lines denote the average values of $I_{c}$,
25 $\mu A$ and 38 $\mu A$, for the frequency range of 70 GHz and
90 GHz, respectively.

\newpage
\begin{figure}[p]
\begin{center}
\leavevmode
\includegraphics[width=0.8\linewidth]{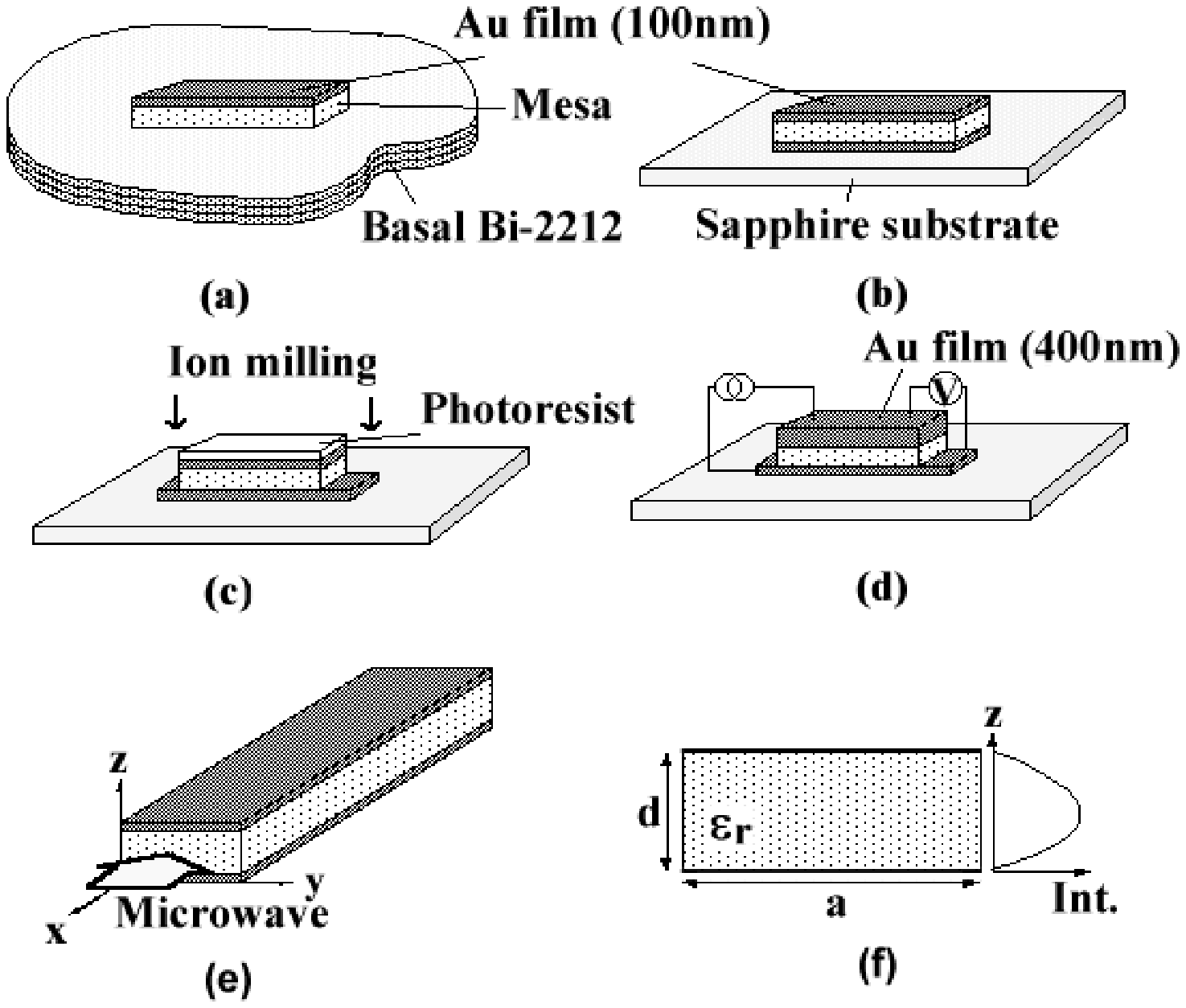}
\caption{}
\end{center}
\end{figure}

\begin{figure}[p]
\begin{center}
\leavevmode
\includegraphics[width=0.5\linewidth]{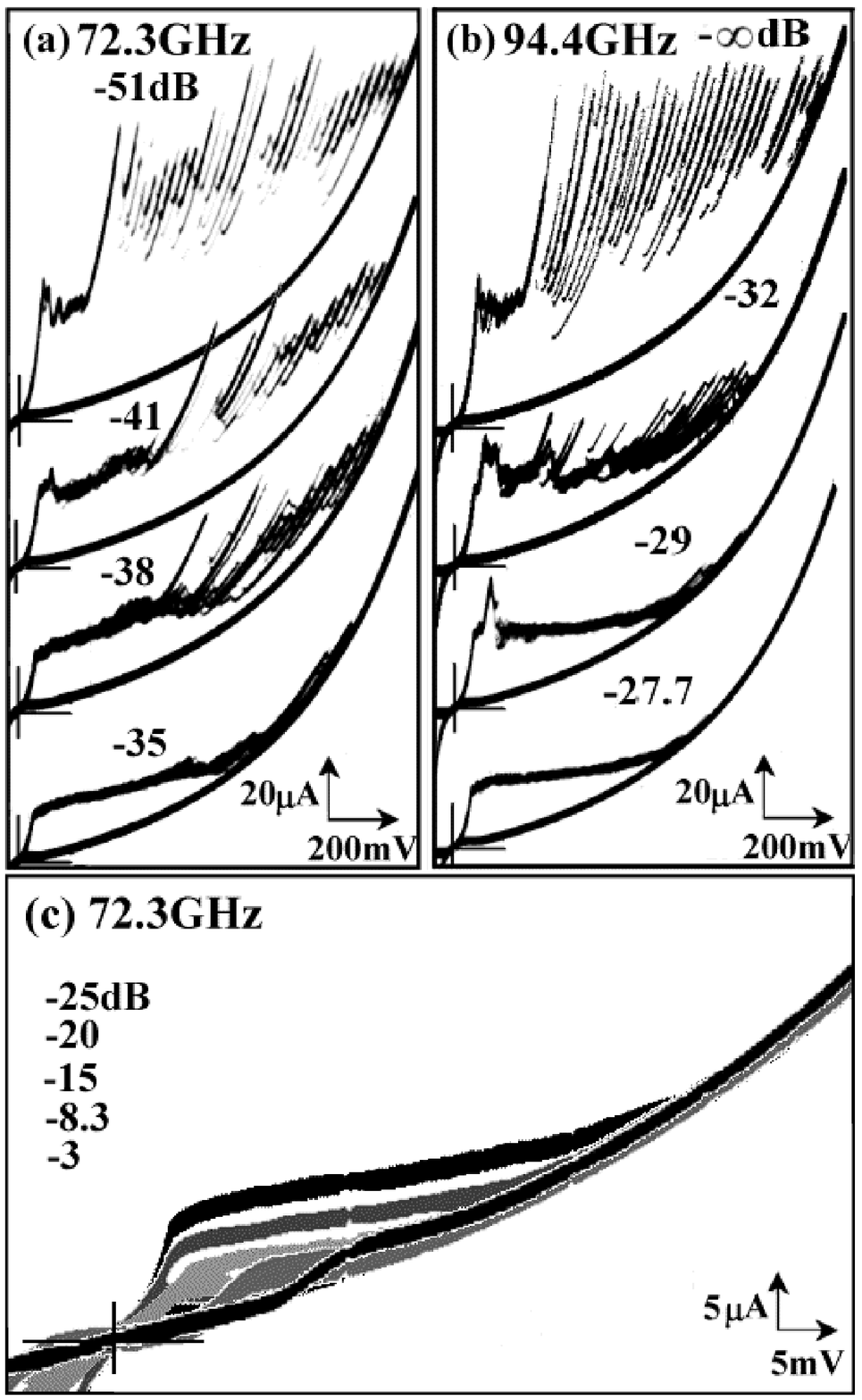}
\caption{}
\end{center}
\end{figure}

\begin{figure}[p]
\begin{center}
\leavevmode
\includegraphics[width=0.7\linewidth]{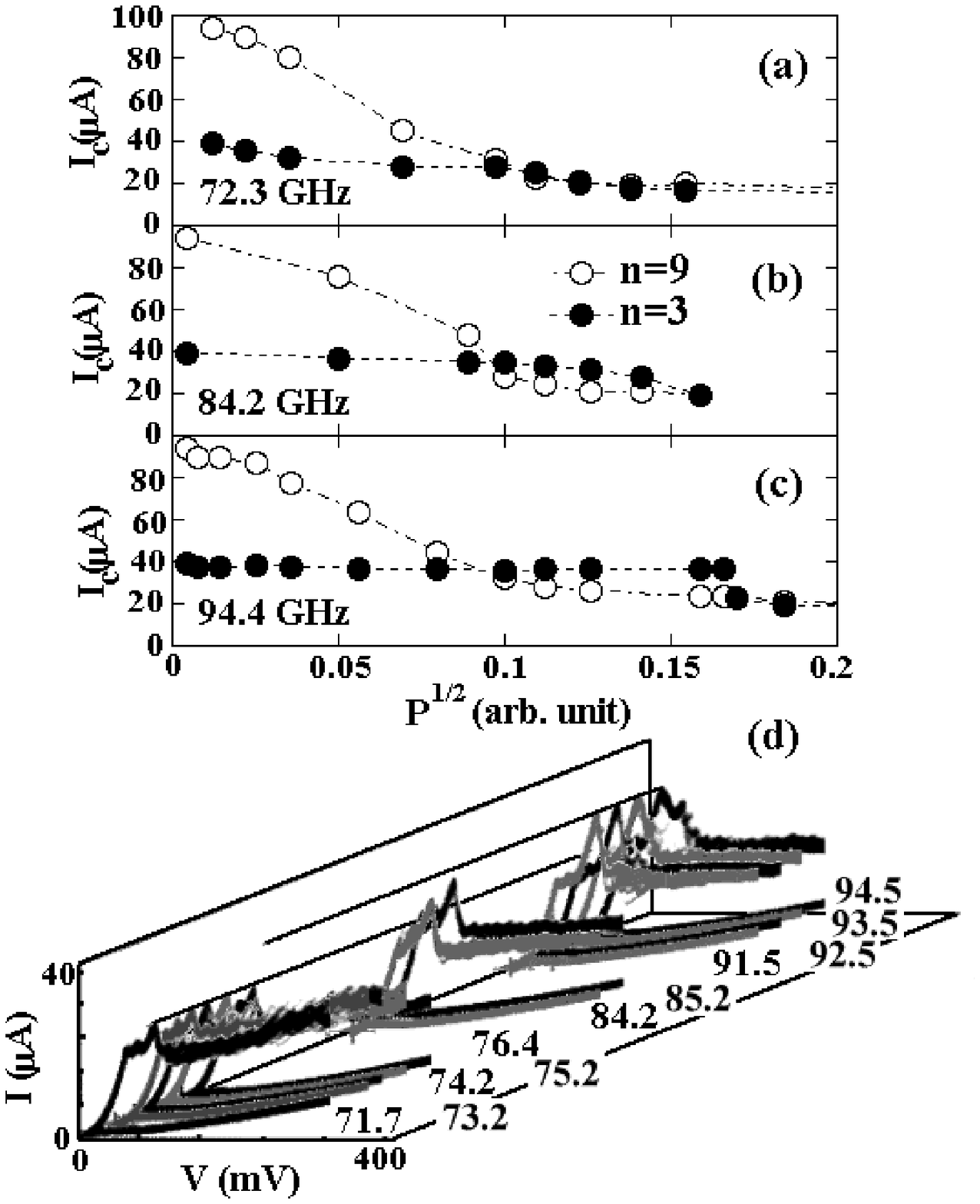}
\caption{}
\end{center}
\end{figure}

\end{document}